\documentclass[12pt,a4paper]{article}
\usepackage{a4wide}
\usepackage{amsmath}
\usepackage{amssymb}
\usepackage{txfonts}
\begin{document} 
{\renewcommand{\thefootnote}{\fnsymbol{footnote}}
\medskip
\begin{center} {\LARGE First Class Constrained Systems and \\ 
\vskip0.4em Twisting of Courant
Algebroids by a Closed 4-form}\\
\vspace{3em}{\Large Markus Hansen$^a$\footnote{e-mail
address: {\tt markus.hansen1-@-gmx.net}} and Thomas
Strobl$^b$\footnote{e-mail address: {\tt
strobl-@-math.univ-lyon1.fr}}}
\\\vspace{2em} $^a$
Friedrich-Schiller-Universit\"at Jena\\
Mathematisches Institut\\
Ernst-Abbe-Platz 2\\
07743 Jena, Germany\\
\vspace{1em}  $^b$
Universit\'e de Lyon, Universit\'e Lyon 1,\\
CNRS UMR 5208, Institut Camille Jordan,\\
43 Boulevard du 11 Novembre 1918,\\
F--69622 Villeurbanne Cedex, France.\\

\vspace{5em}

{\it {\Large In memoriam of Prof.~Wolfgang Kummer}}

\end{center} }

\newcommand{\TS}[1]{{\bf\boldmath \framebox{TS} :#1}}
\newcommand{\MH}[1]{{\bf\boldmath \framebox{MH} :#1}}

\def\2{{\textstyle\frac{1}{2}}}
\def\ba{\begin{eqnarray}}
\def\ea{\end{eqnarray}}
\def\be{\begin{equation}}
\def\ee{\end{equation}}
\def\re{(\ref }
\newcommand{\Kern}{\mathop{\mathrm{ker}}}
\newcommand{\rank}{\mathop{\mathrm{rank}}}
\newcommand{\Ad}{\mathop{\mathrm{Ad}}}
\newcommand{\ad}{\mathop{\mathrm{ad}}\nolimits}
\newcommand{\tr}{\mathop{\mathrm{tr}}}
\newcommand{\Mod}{{\CM}_{\mathrm{cl}}}
\newcommand{\fra}{\mathfrak{a}}
\newcommand{\frg}{\mathfrak{g}}

\let\a=\alpha \let\b=\beta \let\g=\gamma \let\d=\delta
\let\e=\varepsilon \let\ep=\epsilon \let\z=\zeta \let\h=\eta
\let\th=\theta
\let\dh=\vartheta \let\k=\kappa \let\l=\lambda \let\m=\mu
\let\n=\nu \let\x=\xi \let\p=\pi \let\r=\rho \let\s=\sigma
\let\t=\tau \let\o=\omega \let\c=\chi \let\ps=\psi
\let\ph=\varphi \let\Ph=\phi \let\PH=\Phi \let\Ps=\Psi
\let\O=\Omega \let\S=\Sigma \let\P=\Pi
\let\Th=\Theta \let\L=\Lambda \let \G=\Gamma \let\D=\Delta
\def\wtO{\widetilde{\Omega}}

\def\({\left(} \def\){\right)} \def\<{\langle} \def\>{\rangle}
\def\lb{\left\{} \def\rb{\right\}}
\let\lra=\leftrightarrow \let\LRA=\Leftrightarrow
\def\ul{\underline}
\def\wt{\widetilde}
\let\Ra=\Rightarrow \let\ra=\rightarrow
\let\la=\leftarrow \let\La=\Leftarrow

\def\CG{{\cal G}}\def\CN{{\cal N}}\def\CC{{\cal C}}
\def\CL{{\cal L}} \def\CX{{\cal X}} \def\CA{{\cal A}}  
\def\CB{{\cal B}}  \def\CE{{\cal E}}
\def\CF{{\cal F}} \def\CD{{\cal D}} \def\rd{\rm d}
\def\rD{\rm D} \def\CH{{\cal H}} \def\CT{{\cal T}} \def\CM{{\cal M}}
\def\CI{{\cal I}}
\def\CP{{\cal P}} \def\CS{{\cal S}} \def\C{{\cal C}}
\def\CR{{\cal R}}
\def\CO{{\cal O}}
\def\CU{{\cal U}}

\newcommand*{\dR}{{\mathbb R}}
\newcommand*{\dN}{{\mathbb N}}
\newcommand*{\dZ}{{\mathbb Z}}
\newcommand*{\dC}{{\mathbb C}}
\newcommand{\md}{\mathrm{d}}
\newcommand{\mdp}{\mathop{{\mathrm{d}}_{\parallel}}\nolimits}
\newcommand{\mdo}{\mathop{{\mathrm{d}}_{\perp}}\nolimits}
\newcommand{\Diff}{\mbox{\rm Diff}}
\newcommand{\diff}{\mbox{\rm diff}}

\makeatletter
\def\bard{\protect\@bard}
\def\@bard{%
\relax
\bgroup
\def\@tempa{\hbox{\raise.73\ht0
\hbox to0pt{\kern.4\wd0\vrule width.7\wd0
height.1pt depth.1pt\hss}\box0}}%
\mathchoice{\setbox0\hbox{$\displaystyle\mathrm{d}$}\@tempa}%
{\setbox0\hbox{$\textstyle \mathrm{d}$}\@tempa}%
{\setbox0\hbox{$\scriptstyle \mathrm{d}$}\@tempa}%
{\setbox0\hbox{$\scriptscriptstyle \mathrm{d}$}\@tempa}%
\egroup
}
\makeatother

\def\Rboost{\dR}

\def\oG{\hbox{${\cal G}_+^\uparrow$}}

\newtheorem{theo}{Theorem}
\newtheorem{lemma}{Lemma}
\newtheorem{cor}{Corollary}
\newtheorem{defi}{Definition}
\newtheorem{prop}{Proposition}

\newcommand{\proofend}{\raisebox{1.3mm}{%
\fbox{\begin{minipage}[b][0cm][b]{0cm}\end{minipage}}}}
\newenvironment{proof}[1][\hspace{-1mm}]{{\noindent\it Proof #1:}
}{\mbox{}\hfill \proofend\\\mbox{}}
\vskip4em

\begin{abstract}
\noindent 
We show that in analogy
to the introduction of Poisson structures twisted by a closed 3-form
by Park and Klimcik-Strobl, the study of three dimensional sigma
models with Wess-Zumino term leads in a likewise way to twisting of
Courant 
structures by closed 4-forms $H$.

The presentation is kept pedagogical and accessible to physicists as
well as to mathematicians, explaining in detail in particular the
interplay of field transformations in a sigma model with the type of
geometrical structures induced on a target. 
In fact, as we also show, even if one does not know the mathematical
concept of a Courant algebroid, the study of a rather general class of
3-dimensional sigma models \emph{leads} one to that notion by itself.

Courant algebroids became of relevance for mathematical physics lately
from several perspectives---like for example by means of using
generalized complex structures in String Theory. One may expect that
their twisting by the curvature $H$ of some 3-form Ramond-Ramond gauge
field will become of relevance as well. 
\end{abstract}

\setcounter{footnote}{0}

\newpage

\section{Preamble {\rm{{\large{(by T.S.)}}}}} 
It was one of the meritorious  
goals of Prof.~W.~Kummer to promote promising students as soon as
possible in the course of their studies. An important tool in this
context were the three ``Vorbereitungspraktika'' (experimental, but
optionally also theoretical internships of about six weeks length
each), which one performs before tackling the diploma thesis at the
Technical University of Vienna. I made two such internships with him,
one on two-dimensional gravity models, upgraded into a diploma thesis,
and another one on constrained systems. Pursuing further
two-dimensional gravity models within my PhD with Kummer, a hidden
target Poisson structure in such models became transparent in the work
together with P.~Schaller. While this gave rise to the Poisson sigma
model (PSM) and its later twisting by a closed 3-form with C.~Klimcik,
we will consider a topological pendant of the PSM in three spacetime
dimensions, the (as we will define it: also twisted) Courant sigma
model (CSM) that generalizes the Chern Simons theory and can be also
related to gravity models in a spacetime dimension by one more than
two. We think that W.~Kummer would have enjoyed such a development and
extension of previous, in part joint activities (despite his probably
slightly less mathematical or structural interests). In the present
contribution we focus on rather concrete calculations within the
Hamiltonian framework, which Kummer enjoyed very much in the last
decades of his scientific work. We complement this, however, also with
a coordinate-free reinterpretation, which turns out to be less obvious
as one may think at first. Let me mention on this occasion also that I
was always fascinated by the joy Kummer had in calculational projects,
an enthusiasm, that, in the end, was very stimulating for the whole
group surrounding him.

The present account is a report about an internship that I appointed
to a promising student at the FSU Jena some years ago, who is now in
his PhD and is coauthoring this article. Since on the one hand this
enterprise relates in several ways to Kummer's activities and
interests, in particular those tying me with him (cf.~also above), and
on the other hand its result provides the twisting of Courant
algebroids (a lately much discussed mathematical notion, of relevance
in several branches of modern physics and geometry) by closed 4-forms
including the corresponding topological sigma model, we thought it
well adapted as a contribution to a memorial volume for Wolfgang
Kummer. A structurally further-going related analysis about the
appearing current algebra, which would generalize a joint paper with
A.~Alekseev on two-dimensional current algebras to arbitrary
dimensions, may be provided elsewhere.

\section{Introduction}
The Poisson sigma model (PSM) \cite{PSM1,Ikeda}
\be S_{PSM}[X^i,A_i] = \int_\S A_i \wedge \md X^i + \2 \CP^{ij} A_i \wedge
A_j \, .
\label{action1} \ee
has become an important tool within mathematical physics. In the
above, $\S$ is an oriented 2-manifold, $X^i$ and $A_i$ are a
collection of 0-forms and 1-forms on it, respectively, and $\CP^{ij}$
is a matrix depending on the $X$-fields in such a way that it
satisfies a target space Jacobi identity (the brackets denote
antisymmetrization) \be \CP^{l[i}
\CP^{jk]}{}_{,l} = 0 \, .
\label{Jacobi1} \ee The PSM not only comprises a big class of
two-dimensional gravity Yang-Mills gauge theory models
\cite{TK1,Kummerreview,Habil1}, it
also served Kontsevich to find his famous formula \cite{Kontsevich}
for the deformation quantization of Poisson manifolds by means of a
perturbative expansion of its path integral, cf.~\cite{CF1}. Finally,
it is a prototype of a nonlinear gauge theory, which has lead e.g.~to
Lie algebroid extensions of ordinary Yang-Mills theories
(i.e.~non-topological and in any spacetime dimension, cf., e.g.,
\cite{LAYM,StroblMayer}).

For a Hamiltonian formulation one needs to choose $\S$ to contain a
factor $\dR$, corresponding to ``time'' in some sense, and for
simplicity we stick to $\S = S^1 \times \dR$, with a ``periodic''
coordinate $\sigma$ and an evolution parameter $\tau$.\footnote{The
open string Hamiltonian formulation is, albeit slightly more involved,
still very similar. It carries more interesting mathematical
structures, cf.~\cite{CFHam}, but these particular ones are not the
focus of the present consideration.} Plugging $A_i = \lambda_i
\md\tau + p_i \md \s$ into (\ref{action1}), we see that the spatial
components $p_i(\s)$ of the $A$-fields are momenta canonically
conjugate to the ``string'' fields $X^i(\s)$, while its
$\tau$-components $\lambda_i$ serve as Lagrange multipliers for the
following constraints
\be \label{constr1}
G^i(\s) = \partial X^i + \CP^{ij}(X) p_j  \, ,
\ee
where $\partial$ denotes a derivative w.r.t.~$\sigma$ and, on the
r.h.s.~the dependence on $\s$ is understood. By means of the canonical
Poisson brackets and using (\ref{Jacobi1}) one now easily verifies
that the constraints are ``first class'' \cite{Dirac}, meaning that
they close w.r.t.~the Poisson brackets of the field theory (with
structural functions as coefficients), or, in more mathematical terms,
their zero level surface defines a coisotropic submanifold in the
original unconstrained symplectic phase space. Indeed, one finds
\be \{ G^i(\s), G^j(\s') \} = - \delta(\s -\s') \, \CP^{ij}{}_{,k}
G^k(\s) \, ,
\ee
with the structural functions being determined by the $X^k$-derivative
of the $\CP$-tensor, a feature typical in particular for gravity
theories; only in the particular case when $\CP^{ij}(X)$ is linear in
$X$, $\CP^{ij}=C^{ij}{}_{k} X^k$, one reobtains a Yang-Mills type
gauge theory with structure constants given by $C^{ij}{}_{k}$
($S_{PSM}$ reduces to a topological $BF$-theory in this case).

In fact, the consideration can be even reversed: The condition
(\ref{Jacobi1}), which turns $\CP^{ij}$ into a Poisson bivector, is
not only sufficient, it is also necessary for the constraints
$G^i(\s)\approx 0$ to be of the first class. Indeed, it was this
consideration that has lead Schaller-Strobl to find the general
Poisson sigma model after noting corresponding similarities of some
particular two-dimensional gravity or Yang-Mills models---which then
simultaneously turned out to all carry this hidden Poisson target
space geometry. In principle, Poisson geometry could have been
invented by looking at the functional $S_{PSM}$ and requiring it to
define a first class constrained system.

In fact, this strategy was reapplied to disclose a new type of
geometry \cite{HPSM},\footnote{Cf.~also \cite{Park} for another
related, but historically independent consideration.} namely what was
later called twisted Poisson geometry \cite{3Poisson}. Adding a
Wess-Zumino term coming from a closed 3-form $H$ to $S_{PSM}$,
\be S_{HPSM} = S_{PSM} + \int_\S \md^{-1} H \, , \label{HPSM} \ee
which can be interpreted as saying that the symplectic form from
before is changed only by adding a transgression contribution from $H$
to it,
\be \label{eq:om1}
\omega = \oint_{S^1} \! \delta X^i(\s) \wedge \delta
p_{i}(\s)\, \md\s + \2  \oint_{S^1} \!
H_{ijk}(X(\s)) \, \partial X^i
(\s) \, \delta X^j(\s) \wedge
\delta X^k(\s) \, \md\s \; ,
\ee
the constraints of the modified Lagrangian, which still have the form
(\ref{constr1}), are first class, \emph{iff} the following
generalization of (\ref{Jacobi1}) is satisfied:
\be \label{WZPoisson}
\CP^{il} \partial_l \CP^{jk} + {\rm cycl}(ijk)
= \CP^{ii'} \CP^{jj'} \CP^{kk'} H_{i'j'k'} \, .
\ee
In \cite{3Poisson} it was shown that a couple of a bivector and a
closed 3-form satisfying the above condition is in one-to-one
correspondence with $T^*M$-projectable so called Dirac structures
 in split exact Courant algebroids \cite{xxx1,xxx1b,SevLetts}
(cf.~also \cite{DSM} for details).

Courant algebroids became quite fashionable lately within some modern
developments in geometry, like generalized complex structures and pure
spinors (cf., e.g., \cite{Hitchin,Gualtieri}), but also branches
within theoretical physics, such as in String Theory and
supersymmetric sigma models (cf., e.g., \cite{Zabzineetal}). In fact,
in \cite{AlekseevStrobl} it was shown that the Courant bracket appears
naturally within a certain type of current algebra on a phase space
governed by the symplectic form (\ref{eq:om1}). It was moreover found
that maximal systems of first class constraints within this setting
are then in bijection to Dirac structures, which explained also why
the consideration in \cite{HPSM} yielded the Dirac structures as
described in \cite{3Poisson}.

Courant algebroids or Courant structures are the first higher analogue
of Poisson structures: while the latter ones correspond to so-called
NPQ-manifolds of degree one, the former ones are equivalent to
NPQ-manifolds of degree two, cf.~\cite{Roytenberg1} for details on
this. Now, NPQ-manifolds are ideally suited for the construction of
topological field theories following the so-called AKSZ-procedure
\cite{AKSZ}. While for the degree one case one obtains in this way the
PSM (\ref{action1}), cf., e.g.,
\cite{CF-AKSZ}, in the degree two case one obtains \cite{Roytenberg2}
the Courant sigma model (CSM) \cite{Ikeda2,ParkHof}:
\be    S_{CSM}[X^i, A^a,B_i]=
\int_{\S_3}B_i\wedge (\md X^i-\rho_a^i A^a )+\2 \eta_{ab} A^a \wedge \md
A^b + \tfrac{1}{6}C_{abc} A^a \wedge A^b \wedge A^c \, .
\label{action2}
\ee
Here $\S_3$ is a 3-manifold, $X^i$, $A^a$, and $B_i$ are collections
of 0-forms, 1-forms, and 2-forms on it, respectively---where the
number of scalar and 2-form fields is the same and possibly different
to the number of 1-form gauge fields---and $\rho_a^i$, $\eta_{ab}$,
and $C_{abc}$ are structural functions of the Courant algebroid, its
dependence being on the scalar $X$-fields in (\ref{action2}). These
structural functions are to satisfy a sequel of coupled partial
differential equations so as to give rise to the structure of a
general (not necessarily exact) Courant algebroid.

We are not displaying and explaining those equations, the higher
analogue of the equation (\ref{Jacobi1}) above, at this point. Rather,
as we will show in detail, they can be found by applying the same
strategy as the one leading from (\ref{action1}) to (\ref{Jacobi1}),
which be briefly recalled above, but now applied in the context of the
more elaborate action (\ref{action2}). So, without knowing yet what is
a Courant algebroid, its defining conditions can be derived from
requiring the 3-dimensional sigma model above, with a priori
unrestricted structural functions $\rho_a^i$, $\eta_{ab}$, and
$C_{abc}$, to have first class constraints. Moreover, now twisting the
sigma model by a closed 4-form $H$,
\be S_{HCSM} = S_{CSM} + \int_{\S_3} \md^{-1} H \, , \label{action3}
\ee
we will be lead to a higher analogue of the twisting of a Poisson
structure as in (\ref{WZPoisson}), namely the twisting of the
structure of a Courant algebroid by such a 4-form $H$.\footnote{
\label{splitexact}We denote this 4-form again by $H$. It is not to
be confused with the closed 3-form appearing in very particular
Courant algebroids, namely split exact ones. Here we will find a
\emph{generalization} of a Courant structure, in complete analogy
to the generalization of Poisson structures given by
(\ref{WZPoisson}) and induced by the sigma model (\ref{HPSM}).}

In the subsequent section we will first reconsider some general
prototype of (potentially topological) sigma models in two and three
dimensions of $\Sigma$, leading, under relatively mild assumptions, to
(\ref{action1}) and (\ref{action2}), respectively. As a byproduct we
will be able to determine the tensorial character---or the precise
deviation thereof---of the coefficient objects in these two actions,
which will turn out to be particularly essential in the
three-dimensional context. The coefficient function $\eta$ in
(\ref{action2}), for example, will be seen to correspond to a fiber
metric on a vector bundle $E$ that serves as (part of) the target of the
3d sigma model. The $C$-coefficients, on the other hand, are found to
have a highly non-tensorial transformation behaviour
(cf.~Eqs.~(\ref{etatilde}) and (\ref{Ctrafo}) below).

In section \ref{sec:Ham} we perform the explicit Hamiltonian analysis
of the sigma model $S_{HCSM}$ (with yet unspecified structural
functions $\rho$, $\eta$, $C$ and $H$) and determine the necessary
and sufficient conditions on these functions so as to render the
constrained system first class---thus making the sigma model in
particular also topological. These calculations will be performed for
constant $\eta$ (achievable by field redefinitions and corresponding
e.g.~to orthonormal frames in the above mentioned vector bundle $E$), since
this simplifies the basic Poisson brackets and thus the ensuing
calculations considerably. The drawback of this step is that the
structural identities obtained are then known only in orthonormal
frames.

This sounds less restrictive than it in fact \emph{is}: The structural
identities turn out to also contain derivative terms of the fiber
metric $\eta$ and \emph{cannot} be reconstructed from knowing the
structural equations in orthonormal frames only (where these extra
terms vanish identically). It is here where the considerations of
section \ref{sec:Trafo} become essential. However, another related
complication in this context is that the transformation property of
the $C$-coefficients does not correspond to \emph{any} product of
sections of $E$. We will still be able to construct a
(non-$C^\infty$-linear) product on $\Gamma(E)$, the structure
functions of which will agree with the $C$s in orthonormal frames. 

These questions will be dealt with in the final section to this
contribution, putting together the facts from the two sections
\ref{sec:Trafo} and \ref{sec:Ham} before and providing a
coordinate/frame independent or mathematical formulation of what one
may call an $H_4$-twisted (or a Wess-Zumino-) Courant algebroid. It is
given by Definition 1 in section \ref{sec:final}. In this context we
also will take care of providing a \emph{minimal} set of defining
axioms, other structural identities being shown to follow from
them. We conclude with a concrete example of an $H_4$-twisted Courant
algebroid where $H_4$ is exact.

\section{Field redefinitions and
their geometric significance}
\label{sec:Trafo}

\subsection{Two dimensional sigma models without background data on $\S$}
Let us start with the simpler situation
in two dimensions. We first want to address what kind of action
functionals one can construct without any further structure than
orientability of the base manifold $\S$; we do want orientability for
defining the integral. In particular, there will be no metric given on
$\S$, used in most known cases of action functionals already in the
kinetic term of the non-interacting, ``free'' theory---but also
likewisely in the standard type of sigma models, where one uses
metrics on the base or source manifold $\S$ as well as on the target
manifold $M$. We will consider functionals for 0-forms, 1-forms, and
2-forms in two dimensions. (In principle one could also consider local
functionals defined for fields of other tensor type on $\S$, even
without using a metric, but we will not do this here). We will
restrict ourselves to 0- and 1-forms, $(X^i)_{i=1}^{n}$ and
$(A^\a)_{\a=1}^r$, respectively; in two dimensions this restriction is
very mild, however, and we will comment on the small modifications
when considering also 2-form fields at the end of the subsection.

Under these circumstances we are lead to consider functionals of the
following type:
\be S[X^i,A_\a] = \int_\S  e^\a_i(X)\, A_\a \wedge \md X^i + \2 \CP^{\a\b}(X)
\, A_\a \wedge
A_\b  + \2 B_{ij}(X) \, \md X^i \wedge  \md X^j \, ,
\label{action1a} \ee
where the matrices $e$, $\CP$ and $B$ may at this stage
depend arbitrarily on the scalar fields, the latter two being
antisymmetric, certainly. This is the most general ansatz
in the above mentioned context.

We now come to the first type of field transformations, namely
transformations mapping 0-forms into 0-forms only. Being invertible
(and sufficiently smooth) so as to constitute a permitted field
redefinition, clearly this can be interpreted as a coordinate
transformation on the target spanned by the $n$ scalar fields. The
target would be the range of possible values of $X^i$, which, a
priori, would be an $\dR^n$. Using transformations of the just
mentioned type for an eventual gluing, and considering
(\ref{action1a}) as an appropriately understood locally valid
expression only, we can generalize this to considering $X$ as a map
from $\Sigma$ to a general $n$-dimensional (target) manifold $M$.  

As a consequence from this consideration, the last term in
(\ref{action1a}) receives the interpretation of the pullback to
$\Sigma$ by $X$ of a 2-form $B$ on $M$.\footnote{Note that here certainly $B$
is not a field but a fixed 2-form on $M$, which only encodes part of
the kinetic and interaction terms for the scalar fields. This is as in
String theory, where one denotes such a term by precisely the same
symbols conventionally, and where then $B$ becomes a 2-form field only
on the \emph{target} by means of a dynamics induced implicitly by
string fluctuations.} To also give a geometric meaning to the other
quantities in the above action, we consider transformations of the
form $A_\a \mapsto M^\b_\a(X) A_\b$ (for invertible, smooth matrices
$M$). Since there are no derivatives acting on the $A$-fields, they
just imply a tensorial transformation property of the $\a$-indices in
$e$ and $\CP$. In particular, we may conclude that $A_\a$, besides being
1-forms on $\Sigma$, corresponds to components (indexed by $\a$) of
sections in some rank $r$ vector bundle $E$ living over $M$ ($M^\b_\a$
corresponding to local frame changes in this bundle, moreover). This
implies then that $\CP \in \Gamma(\Lambda^2 E^*)$ and that $e \in
\Gamma(E^* \otimes T^*M)$, where $E^*$ denotes the bundle dual to $E$;
$e$ can equivalently be viewed as a map from $E$ to $T^*M$.  

Let us now, as the main restriction in this context, assume that this
map $e$ provides an isomorphism, $e \colon E
\stackrel{\sim}{\rightarrow} T^*M$, which in particular implies that
the number $n$ of 0-form fields and the number $r$ of 1-form fields need
to be equal and that the then $n \times n$ matrices $e$ are everywhere invertible.
In fact, under this condition, $e$ is seen to be nothing but a
\emph{vielbein} on $M$, and $A_\a$ then turns out to be the
components of a 1-form in $M$ in a potentially non-holonomic basis,
while $\CP$ becomes a bivector field: $E\cong T^*M$ implies $E^* \cong
TM$, i.e.~in a holonomic basis $\partial_i$ of $TM$ one has $\CP = \2
\CP^{ij} \partial_i \wedge \partial_j \in \Gamma(\Lambda^2TM)$. A
field redefinition of $A_a$ of the form $A_i := e^\a_i (X) A_\a$,
which induces a redefinition of the coefficient matrix in the
quadratic $A$-term $e_\a^i e_\b^j \CP^{\a\b} =: \CP^{ij}$, where
$e_\a^i$ is the inverse vielbein, is now seen to just correspond to a
change from a general frame to a holonomic basis of $TM$. The action
(\ref{action1a}) is now seen to be identical to (\ref{HPSM}) with
$H=\md B$ after these change of variables.

Diffeomorphisms of the target, $X^i \mapsto \widetilde{X}^i(X)$, can
now be compensated directly with a corresponding redefinition of the
$A$-fields in the holonomic frame, $A_i \mapsto \widetilde{A}_i =
\frac{\partial X^j}{\partial \widetilde{X}_i}A_j$. Certainly this is
in general \emph{not} a symmetry of the action functional, since the
explicit form of the matrices $\CP^{ij}$ and $B_{ij}$ as functions of
$X$ will change---except if the generating vector field of the
diffeomorphism Lie annihilates the bivector field and the 2-form, in
which case one has a rigid symmetry giving rise to Noether
charges.\footnote{The analogue of this in String theory with a
Minkowski target are momentum and angular momentum, as the Noether
charges of the Poincare isometry group.}

There are further field redefinitions of less immediate geometric
significance. One of these corresponds to a shift of the A-fields by
terms proportional to $\md X$. Such transformations are easily seen to
change the $B$-contribution to (\ref{action1a}) and they can be shown
to even permit to get rid of this contribution altogether; this is by
far less immediate and was in fact proven rigorously only for small
enough $B$ in \cite{DSM} (but cf.~also \cite{Izawa}). Assuming this to
hold true also for general $B$, it implies that only the deRham
cohomology class of $H$, entering as a Wess-Zumino term in
(\cite{HPSM}), has a physical significance, at least if no additional
meaning is attributed to distinguished fields or target coordinates in
the action (like it might happen in some particular gravitational
applications, for example). The geometrical significance of these
changes of $H$ by an exact term $\md B$ is less immediate as well:
Note, for example, that this permits to change a Poisson tensor $\CP$
into one that is only $\md B$-twisted Poisson. Still, there is some
geometrical notion behind this, which, interestingly, relates in a
different way again to Courant algebroids, the main subject of this
article, in their interplay with sigma models: (possibly twisted)
Poisson structures are particular so-called Dirac structures, a
particular type of subbundles in exact Courant algebroids. As a bundle
an exact Courant algebroid is isomorphic to $T^*M \oplus TM$, where
the isomorphism corresponds to a splitting in an exact sequence and
changes of this splitting correspond precisely to some $B\in
\Omega^2(M)$ as above.\footnote{The above $E=T^*M$ enters this picture
in so far as the Dirac structure corresponding to a bivector
field or to a twisted Poisson structure provides by itself an
isomorphism of $T^*M$ into an appropriate subbundle of $T^*M \oplus
TM$, so that, after the choice of a splitting, $E$ can be identified
with this subbundle of the exact Courant algebroid.} We refer to \cite{DSM}
for further details.

We conclude this subsection with some remarks on possible
generalizations. The main assumption leading to an identification of
(\ref{action1a}) with (\ref{HPSM}) (for exact $H$) resulted from
requiring $e \colon E \to T^*M$ to be an isomorphism. Even if the
number $n$ of scalar fields and the rank $r$ of $E$ are equal, $e$
might still have a kernel, for example. In fact, if one permits such a
kernel, one is lead to a somewhat more general sigma model than one of
the form of the twisted PSM (\ref{HPSM}), namely one that is of the
form of a so-called Dirac sigma model \cite{DSM}---more precisely, to
the part of it that was called topological there for not depending on
additional background data like a metric on $\S$ (cf., e.g.,
eqs.~(17-20) and eq.~(24) in \cite{DSM}). The restriction to $r=n$, on
the other hand, seems less restrictive than one might believe at first
sight. If $r < n$, it corresponds to $r=n$ with $e$ having a kernel of
dimension $n-r$ and correspondingly many $A$-fields not entering the
action at all.
If, on the other hand, $r>n$, one should be
able to eliminate excess $A$-fields (at least up to potential global
issues): namely those components in the kernel of $e$ enter the action
at most quadratically and only algebraically and then can be
correspondingly eliminated with their own field equations. Suppose, for
example, that $A_1$ and $A_2$ are not present in the $A \wedge \md X$
part of the action and that they enter (\ref{action1a}) only via
$\int_\S A_1\wedge A_2$. Thus, variation w.r.t.~these two fields
require them to vanish. Correspondingly, this term, and thus any
$A_1$- and $A_2$-dependence in this example can be dropped without
changing the physical content of the functional at all.

Finally we briefly comment on not considering also 2-form fields in
the present context. In fact, in the spirit of this section, any 2-form
field can enter an action as (\ref{action1a}) only linearly, then
being multiplied with some function $f(X)$. Variation w.r.t.~this
field yields a constraint $f(X)=0$ which, in the smooth case, singles
out a submanifold of $M$. The sigma model with such a 2-form field, or
several of them, then just reduces effectively to one without those
fields but defined on a smaller target, namely the one of the original
$M$ where the respective functions vanish. The situation can become
more interesting, certainly, if the subspaces singled out by the
vanishing of functions are singular and not just submanifolds. The
explicit conditions on a PSM-type functional in the presence of such
2-form additions to be topological were studied in
\cite{BatalinMarnelius}.

An action functional of the type (\ref{action1a}) to have a maximal
number of possible gauge symmetries and to not carry any propagating
degrees of freedom poses certain conditions on the tensors on the
target of $M$, which are most efficiently found in the Hamiltonian
framework. In the case of (\ref{HPSM}) this lead to (\ref{WZPoisson}),
for example. Although the absence of propagating degrees of freedom
together with the absence of any background structure used for the
definition of such a functional is sufficient to get topological sigma
models, it is not always necessary. An example of a topological sigma
model which uses a background metric on $\S$ (as well as a metric on
$M$) is the G/G WZW model (cf., e.g., \cite{G/G}), or, more generally,
the (full) Dirac sigma model \cite{DSM}. The presence of such
auxiliary structures can be used also as another argument for
restriction to 0- and 1-forms in two dimensions: any 2-form is Hodge
dual to a 0-form. This argument can be used, however, only in this
\emph{extended} context for a convincing exclusion of 2-form fields,
where, on the other hand, there then are also uncountably more
possibilities for the construction of an action functional out of 0-
and 1-form fields than those parametrized in (\ref{action1a}).

\subsection{Three dimensional sigma models}
\label{3dsigma}
We now turn to sigma models that can be defined without any background
structures on an orientable three dimensional base manifold $\S$. In
analogy to before we consider functionals for 0-form fields
$(X^i)_{i=1}^{n}$, 1-form fields $(A^a)_{a=1}^r$, and now also 2-form
fields $(B_\a)_{\a=1}^s$---the omission of top degree-form fields again poses
essentially no restriction. A most general ansatz in this context
takes the following form  
\ba    S[X^i, A^a,B_\a]&=&
\int_{\S} e^\a_i\, B_\a \wedge \md X^i -\rho_a^\a \,B_\a \wedge A^a +
\2 \eta_{ab}\, A^a \wedge \md A^b + \tfrac{1}{6}C_{abc}\, A^a \wedge A^b
\wedge A^c \nonumber \\ &&
\!\!\!\!\! + \: \2 \Lambda_{aij}\, A^a \wedge \md X^i \wedge \md X^j +
\2 \Delta_{abi} A^a \wedge A^b \wedge \md X^i +
\tfrac{1}{6}
F_{ijk}\, \md X^i \wedge \md X^j \wedge \md X^k
\label{action2a}
\ea
where $e^\a_i$, $\rho_a^\a$, $\eta_{ab}$, $C_{abc}$, $\Lambda_{aij}$,
and $F_{ijk}$ are functions of $X$, parametrizing the action
functional.They have the obvious symmetry properties like
e.g.~$C_{abc}$ being completely antisymmetric or $\eta_{ab}$ being
symmetric in the exchange of indices.

In analogy to before we restrict ourselves to the case of $e^\a_i$
being invertible. In addition, here we also require the likewise
coefficient matrix $\eta_{ab}$ nondegenerate as well. One may expect
that relaxing one or the other of these conditions can lead to
interesting generalizations---for example, in the two-dimensional
setting this step permits the more general also topological Dirac
sigma model---, but we will not pursue this here further. Instead, we
will now make use of the non-degeneracy of $e$ to again simplify the
above action by means of appropriate field redefinitions.

First we introduce $B_i :=  e^\a_i\, B_\a$. This one can always do,
certainly, but only in the invertible case we can use $B_i$ as new
fields, by introducing $\rho_a^i := \rho_a^\a e_\a^i$, where, as
before, $e_\a^i$ can be regarded as inverse vielbein. In addition to
replacing $e$ by a unit matrix when it is invertible, field
redefinitions also permit to put $\eta_{ab}$ into constant normal form
\emph{and} to get rid of the terms with coefficient $\Lambda$ and
$\Delta$ altogether. Clearly, redefining $B_i$ by $B_i - \2
\Lambda_{aji}\, A^a \wedge \md X^j$, we eliminate the $\Lambda$-term while
simultaneously we only have to change the coefficient
$\Delta_{abi}$ to $\Delta_{abi}^{\rm{new}} := \Delta_{abi}+\rho^j_a
\Lambda_{bij}$. Similarly we can now get rid of the $\Delta$-term by a
subsequent shift $B_i \mapsto B_i - \2
\Delta^{\rm{new}}_{abi} A^a \wedge A^b$, which now only changes the
coefficient of the cubic $A$-term to $ C^{\rm{new}}_{abc}=C_{abc} + 3 \rho_{[a}{}^i
\Delta_{bc]i}$, where the brackets $[ \ldots ]$ denote
antisymmetrization of the indices enclosed. In this manner we
brought the above action already into the form (\ref{action3}) with
$H=\md F$. We are thus left with the analysis of this action
further on.\footnote{Depending on the context, we will consider this
action---or likewisely (\ref{action2})---as the one of the (twisted)
Courant sigma model or just as a sigma model of this \emph{form} with
structural functions not yet fulfilling the identities needed to
correspond to a (twisted) Courant algebroid.}

We are now left with analysing the field transformations of more
immediate geometrical significance. First of all there are again the
diffeomorphisms of the target of the sigma model, certainly, which
determine also the tensorial character of the index $i$ in $\rho^i_a$
as well as that $F \in \Omega^3(M)$, as anticipated already in the
identification $H=\md F$ mentioned above. The diffeomorphisms induce
certainly a likewise transformation of $B_i$, while not effecting the
$A$-fields. The latter 1-form fields take again values in some rank
$r$ vector bundle $E \to M$ (more precisely, $A \in
\Omega^1(M,X^*E)$). We are left with analysing changes of quantities
induced by transformations $A^a \mapsto \widetilde{A}^a$,   
\be A^a  = M^a_b(X) \widetilde{A}^b \, , \label{frames} \ee
corresponding to changes of local frames in $E$. Obviously the
Wess-Zumino term in (\ref{action3}), stemming from some $H
\in\Omega_{\rm{closed}}^4(M)$, is not effected by such transformations
and we can focus on (\ref{action2}) for this purpose.

Note that at this point we have not yet put $\eta$ into some normal
form, since this would restrict the permitted local frames in $E$ to
orthogonal ones w.r.t.~$\eta$ viewed as a fiber metric on $E$. This
will be an important issue since, as we will see in the end of the
analysis within the present article, on the one hand many
\emph{different} geometrical quantities within the present setting
will be seen to coincide in orthonormal frames (and in their index
representation differ decisively from one another by derivatives of
$\eta_{ab}$ only), while, on the other hand, the Hamiltonian analysis
simplifies drastically in orthonormal frames so that, at least for
that purpose, we \emph{do} want to restrain $\eta$ to a constant
normal form. But to be able to retrieve other involved objects in
general frames again from there, we need to know their transformation
properties w.r.t.~a general transformation as in (\ref{frames}). We
now first observe that a transformation of the form (\ref{frames})
induces also a nontrivial $\Delta$-contribution to the action, namely
one with $\Delta_{abi} = 
-\tfrac{1}{2} \widetilde{\eta}_{ab,i}$\footnote{Note that this
contribution is no more present if $\eta_{ab}$ was already put to
constants and if one restrains $M^a_b$ to respect that, i.e.~to
correspond to orthogonal transformations. --- In this article, we
use the convention that $f,_i$ denotes the partial derivative of a
function $f$ w.r.t.~$X^i$.}, where
\be \label{etatilde}  \widetilde{\eta}_{ab} = M^c_aM^d_b \eta_{cd} \ee
denotes the components of $\eta$ in the new frame. (It is also this
equation, together with the required non-degeneracy, that justifies to
regard $\eta$ as a fiber metric on $E$).  To get rid of the unwanted
contribution in the action, we learnt above that we can do this by
accompanying (\ref{frames}) by a $B$-field transformation, $B_i \to
\widetilde{B}_i$ where
\be B_i = \widetilde{B}_i + \tfrac{1}{4}
\widetilde{\eta}_{ab,i} \widetilde{A}^a \wedge \widetilde{A}^b \, .\ee
This, on the other hand, by itself leads to a new additive
contribution to the coefficient of the cubic $A$-term, thus rendering
$C_{abc}$ to have a \emph{non-tensorial} transformation property;
besides the obvious $\widetilde\rho^i_b= \rho^i_a M^a_b$ one finds
\be \widetilde C_{abc}=M^d_a M^e_b M^f_c C_{def}
-3M_{[a}^d M_b^e M^f_{c],d}\eta_{ef} \, , \label{Ctrafo}
    \ee
where from now on we use the further on useful abbreviated notation
\be f_{,a} \equiv \rho^i_a f_{,i} \label{kommaa} \ee
for derivatives along letters of the beginning of the alphabet. So,
while thus $\eta$ is seen to correspond to a fiber metric on $E$,
which, by an appropriate choice of $M_a^b$ we can always put to some
constant normal form and $\rho$ is found to be an element of
$\Gamma(E^* \otimes TM)$, or, equivalently, a vector bundle map
\be \rho \colon E \to TM \, , \label{rho} \ee
the differential geometric meaning of $C$ is much more intricate. We
will clarify its meaning after having derived the equations the
structural functions have to satisfy in an orthonormal basis in the
subsequent section so as to render (\ref{action3}) topological.
 
Finally, we remark that also in the three dimensional context it is
only the non-exact WZ-term that gives something qualitatively
new. Here in three dimensions this is even relatively easy to see
explicitly in the sigma model: A transformation $B_i \mapsto B_i +
\tfrac{1}{6} F_{jki} \md X^j \wedge \md X^k$ adds a term of the form
of the last one in (\ref{action2a}). We also produce a nontrivial
$\Lambda$-term in this manner, but we already know how to remove it
by further $B$-field transformations. The upshot is that such a
combined transformation only changes $H$ to $H + \md F$ in
(\ref{action3}). Indeed, the situation is very analogous to the
geometry one finds from the two-dimensional sigma model, it is a
higher analogue of it in several ways and we will display here only
parts of the full story.  

\section{Hamiltonian analysis}
\label{sec:Ham}

\subsection{Hamiltonian formulation}
In this section we perform a Hamiltonian analysis of the action
(\ref{action3}). For this purpose we choose $\S_3 = \S \times \dR$
with $\S$ an oriented, compact 2-surface without boundary, and in a
first step we only regard the usual local part of the action given by
(\ref{action2}). Since the ensuing Hamiltonian formulation is much
easier when $\h_{ab}$ is constant, we will assume this to be the case
within this section. Also, since $\eta$ is nondegenerate, we can use 
it also freely to raise and lower letters from the beginning of the
alphabet, so, e.g.,
\be \r^{ai}(X) = \h^{ab}\r^i_b(X) \, , \label{b}
\ee
and, since $\h$ is constant in the given frame, this can be done also
with quantities that are hit by derivatives. In the above, $\h$ with
upper indices denotes, as usual, the inverse to $\h$ with lower
indices (agreeing, at the same time, with one of the two having
changed both index positions by the respective other one in the
indicated way---a feature where the symmetry of $\h$ is essential
for). As mentioned repeatedly already, except for appropriate
smoothness conditions, at this point we do require nothing more of the
coefficient functions of the $X$-fields in (\ref{action2}).

In fact, the action (\ref{action2}) is already in a Hamiltonian
form. To see this we decompose the forms appropriately: $A^a = \CA^a +
\Lambda^a \rd \tau$, where $\CA^a$ are 1-forms on $\S$ at a fixed
value of the evolution parameter $\tau$, and $\Lambda^a$ likewisely
0-forms. Analogously, we have $B_i = p_i + \rd
\tau \wedge \lambda_i$, with 2-forms and 1-forms $p_i$ and $\lambda_i$, respectively.
Plugging this decomposition into  (\ref{action2}), using $\rd = \rd_\S
+ \rd \tau \wedge \partial_\tau$ and denoting the $\tau$-derivative of a
quantity by an overdot, $ \partial_\tau \phi \equiv \dot{\phi}$,
we find
\be     S_{CSM}=
\int_\dR \left[ \int_{\S}p_i \dot{X}^i - \2 \eta_{ab} \CA^a \wedge
\dot{\CA^b} + \lambda_i \wedge G^i + \L_a H^a
\right] \wedge
\rd \tau \, ,
\label{actionHam}
\ee
with $G^i \equiv \md_\S X^i-\r^i_a\CA^a$ and $H^a\equiv \md_\S
\CA^a+\2 C^a_{bc}\CA^b\wedge \CA^c-\r^{ai}p_i$. The first two terms
are a symplectic potential; such a potential gives rise to a
symplectic form by replacing $\tau$-derivatives by differentials of
the respective field (in field space, we will denote the corresponding
exterior derivative by $\delta$ for clarity, as we did already in
(\ref{eq:om1})) and taking the negative exterior derivative of the
result.\footnote{The pioneering work about a Hamiltonian formulation
of gauge theories goes back to Dirac \cite{Dirac}. For a somewhat
simplified version, applicable also in the present context, cf.,
e.g.,\cite{FaddeevJackiw}.} Denoting the field $\CA^a$ again by simply
$A^a$, this evidently yields
\be
\o_{CSM} = \oint_\S\d X^i\wedge\d p_i
     +\2\oint_\S\h_{ab}\d A^a\wedge\d A^b  \, .
\ee
The remaining two terms in (\ref{actionHam}) give rise to constraints
only, with the  $\lambda_i$ and $\Lambda_a$ being their Lagrange
multiplier fields. Thus, in the simplified notation, the following
currents have to vanish:
\ba
    G^i(\s) &=&\md X^i-\r^i_a(X)A^a \label{G}\\ \label{J}
    J^a(\s) &=&\md A^a+\2 C^a_{bc}(X)A^b\wedge A^c-\r^{ai}(X)p_i.
\ea
Here $X^i$, $A^a$, and $p_i$ are now functions, 1-forms, and 2-forms
on the 2-surface $\S$, respectively, and, correspondingly, also the
suffix $\S$ has been dropped on the deRham differential. For some
purposes it is useful to introduce test objects so as to obtain true
functions on the field theoretic phase space. Let $\mu_i$ and
$\varphi_a$ be such a collection of test 1-forms and 0-forms on $\S$,
respectively, and set:\footnote{For simplicity, we consider a fixed
frame and do not permit any of the test objects to depend on the
$X$-field (or any other field) in what follows. --- The integration
symbol $\oint$ has been chosen, here and already before, so as to
stress that there are no boundary contributions to the integral due to
the choice of $\S$.}
\ba
    G[\mu] &:=& \oint_\S \mu_i \wedge G^i  \label{G2} \\
    J[\varphi] &:=& \oint_\S \varphi_a J^a  \label{J2}
\ea
These functions on phase space $\CM$ have to vanish for all choices of
test objects (which can be considered as generalized labels for the
constraints), which defines the constraint surface $\CC\subset \CM$;
this is a consequence following from the action functional $S_{CSM}$.

We will in the following require that in addition also mutual Poisson
brackets of the constraints vanish on $\CC$, which is a restriction on
the structural functions in the action. In the nomenclature of Dirac
this is denoted as
\be G[\mu] \stackrel{!}{\approx} 0  \stackrel{!}{\approx} J[\varphi]
\qquad \forall \mu_i, \varphi_a  \, .  \label{constraintsurface}
\ee
It means that also the Hamiltonian vector fields of the constraints,
restricted to the constraint surface $\CC$, are required to be tangent
to it. In a more mathematical language the first class property is
tantamount to saying that the (here infinite dimensional) submanifold
$\CC$ of the original (here weakly symplectic) phase space manifold
$\CM$ is coisotropic.\footnote{Cf., e.g., \cite{BojowaldStrobl0112074}
for several equivalent characterizations of this notion within the
finite dimensional setting.}

Twisting the sigma model by a closed 4-form as in (\ref{action3}),
gives a contribution to the symplectic form only. In fact, the action
(\ref{action3}) is uniquely valued only when $H=\rd h$ is exact, in
which case it amounts to adding the pullback of $h\in \O^3(M)$ by the
map $X \colon \S_3 \to M$ to the action (\ref{action2}). In this case,
(\ref{action3}) with $\S_3 = \dR \times \S$ is understood to be
\be S_{HCSM}=S_{CSM}+\int_\dR\left[\int_\S\2 h_{ijk}d_\S X^i\wedge
d_\S X^j\dot X^k\right]\wedge d\tau,\ee so that the new term clearly
gives a contribution to the symplectic potential only. The
corresponding contribution to the symplectic form depends on
$\rd h$ only, 
\be \o_{HCSM}=\o_{CSM}+\2\oint_\S (h_{jil,k}+h_{kjl,i})\d X^k\wedge\d X^l\wedge
d_\S X^i\wedge d_\S X^j \, ,\ee
and can be defined for arbitrary closed $H$: 
\be \o_{HCSM} =  \oint_\S\d X^i\wedge\d p_i +
\tfrac{1}{4}\oint_\S H_{ijkl}\d X^i\wedge\d X^j
\wedge\rd X^k\wedge\rd X^l
     +\2\oint_\S\h_{ab}\d A^a\wedge\d A^b \, . \label{symp}
\ee
This form remains (weakly) nondegenerate for any choice of $H$;
closedness of $H$ (on the target $M$) becomes necessary for the
closedness of the symplectic form $\o_{HCSM}$ on the field theoretic
phase space $\CM$. In the Hamiltonian formulation a Wess-Zumino term
can be added without any integrability condition on the closed
$d+1$-form; this would arise upon geometric prequantization, for
example.

\subsection{Constraint algebra}
To calculate Poisson brackets among the constraints, we first
display the elementary Poisson brackets as they follow from the
symplectic form (\ref{symp}). By standard methods one obtains, written
in components,
\begin{align}
    \lb X^i(\s),\tilde{p}_j(\widetilde\s)
\rb &= \d^i{}_j\, \d(\s-\widetilde\s),\\
    \lb \tilde{p}_i(\s),\tilde{p}_j(\widetilde\s)\rb &=
    \2 H_{ijkl}X^k{}_{,\mu}X^l{}_{,\nu}
\, \e(\mu\nu)\, \d(\s-\widetilde\s),\\
    \lb A^a_\mu(\s),A^b_\nu(\widetilde\s)\rb &= 2\h^{ab}\,
    \e(\mu\nu)\,
\d(\s-\widetilde\s),
\end{align}
with the other brackets vanishing. Here $A^a=A^a_\mu d\s^\mu$,
$p_i=\tilde{p}_i \,\rd\s^1\wedge \rd\s^2$, $\delta(\sigma -
\widetilde\s)$ is the delta function w.r.t.~the measure $\rd\s^1\wedge
\rd\s^2$, $\e(\mu \nu)$ denotes the $\e$-symbol normalized according
to $\e(12)=1$, and quantities on the r.h.s.~are understood to depend
on either $\s$ or $\widetilde\s$. Using again test objects,
\be \label{elem}    \hat{X}[\alpha]:=\int_\S\a_i X^i,
        \qquad \hat{A}[\mu]:=\int_\S\mu_a\wedge A^a,
        \qquad \hat{P}[\varphi]:=\int_\S\varphi^i p_i,
\ee
where $\varphi^i$, $\mu_a$, and $\a_i$ are 0-, 1-, and 2-forms on
$\S$ respectively, this can be rewritten as
\begin{align}
    \lb \hat{X}[\a],\hat{P}[\varphi]\rb &=
\int_\S \varphi^i \a_i,\\
    \lb  \hat{P}[\varphi] , \hat{P}[\tilde{\varphi}]\rb &=
    \2\int_\S \varphi^i\tilde{\varphi}^jH_{ijkl}
    dX^k\wedge dX^l,\\    \lb  \hat{A}[\mu], \hat{A}[\tilde{\mu}]\rb
&= 2\int_\S\mu_a \wedge\tilde{\mu}^a,
\end{align}
all other brackets between the elements (\ref{elem}) vanishing.

Now we are ready for the real calculation. Using the above elementary
brackets, one computes those between the constraints (\ref{G2}),
(\ref{J2}). We display here only the result of the somewhat lengthy
calculation. One obtains:
\be
    \Bigl\{G[\mu],G[\tilde\mu]\Bigr\}
        =\int_\S\rho^{ai}\rho^j_a
\, \mu_i\wedge\tilde{\mu}_j  \label{GG}
\ee
\be
    \Bigl\{G[\mu],J[\ph]\Bigr\}
        =\int_\S G^j \wedge\mu_i  \ph^a\, \rho^i_{a,j}
            +A^b \wedge \ph^a\mu_i\Bigl(\rho^{ci} C_{abc} 
            +2\rho^i_{[a,b]}\Bigr)  \label{GJ}
\ee
\be
    \begin{split} \label{JJ}
        \Bigl\{J[\ph],J[\tilde\ph]\Bigr\}=
            &-\int_\S\Bigl(J^fC^{d}_{ef}+G^i\wedge A^fC^{d}_{ef,i}\Bigr)\ph^e\tilde\ph_d\\
            &+\int_\S\2\bigl(G^k\wedge G^l+2
G^{[k}\rho^{l]}_c\wedge A^c\bigr) \ph_a\tilde\ph_d\rho^{ai}\rho^{dj}H_{ijkl}
               \\
 &-\int_\S\ph^b\tilde\ph^a p_i
                \Bigl(\rho^{ci}C_{abc} +2\rho^i_{[a,b]}\Bigr)\\
            &-\int_\S\ph^a d\tilde\ph^b \wedge A^c\Bigl(C_{abc}+C_{bac}\Bigr)\\
           &+\int_\S\ph^a\tilde\ph^bA^c\wedge A^d\Bigl(\tfrac{1}{2}C_{bae}C^e_{cd}-C_{bad,c}
                -C_{a]cd,[b}+
C_{aec}
C_{b}{}^e{}_d
+\2\rho_a^{i}\rho_b^{j}\rho^k_c\rho^l_d 
H_{ijkl}\Bigr).
    \end{split}
\ee
We used the convention $\eta_{ab}C^b_{cd}=C_{acd}$ here (cf.~also
eqs.~(\ref{kommaa}) and (\ref{b}), so that e.g.~$C_{abc,d}\equiv
C_{abc,i}\rho^i_d$).  Now we can determine the necessary and
sufficient conditions for the constraints to be first class. 

Note in this context that the test objects can be chosen
arbitrarily. In particular then the vanishing of (\ref{GG}) implies by
a standard argument (the test objects being arbitrary) that 
\be  \rho^i_a\rho^j_b\eta^{ab}=0 \, . \label{rhorho}
\ee
We remark in parenthesis that this certainly has to hold for any point
in $M$ since any such a point can be image of the map $X \colon \S \to
M$. 

Next we regard (\ref{GJ}), to vanish on (\ref{constraintsurface}),
which in particular implies that the first term on the right hand side
of eq.~(\ref{GJ}) is zero on this surface. There is now one
qualitatively more complicated step than the one in the 1+1
dimensional context of the Poisson sigma model. There the constraints
were 1-forms and on the spatial slice $S^1$ there are no integrability
conditions. Here, there are no integrability conditions for the 2-form
constraints $J=0$, $\S$ being two-dimensional, whereas applying the
deRham differential $\d$ to the 1-form constraints $G=0$, leads, upon
usage of these two equations (\ref{constraintsurface}), to
\be  \tfrac{1}{2}\bigl(\rho^{ci}C_{cab}+2
	\rho^i_{[a,b]}\bigr)A^a\wedge A^b + \rho^{aj}\rho^i_a p_j = 0
	\, .  \label{integrability}
\ee 
The second term was found to necessarily vanish in 
eq.~(\ref{rhorho}) above. We want to conclude from (\ref{GJ}) that 
$\rho^{ci} C_{abc} +2\rho^i_{[a,b]}=0$, which, using that $C_{abc}$ is
completely antisymmetric, can be rewritten also as 
\be  \rho^j_{a}\rho^i_{b,j}- \rho^j_{b}\rho^i_{a,j} =C^c_{ab} \rho^i_c
\, . \label{rhomorph}
\ee 
It is, however, precisely this equation that also enters the
integrability condition (\ref{integrability}) and we want to make sure
to avoid circular reasoning. We need to choose $A^a$ at a given point
on $\Sigma$ sufficiently general to conclude (\ref{rhomorph}) from the
restriction of (\ref{GJ}) to (\ref{constraintsurface}). The main
difficulty at this point is that even at a given point $p$ on $\S$ the 
1-forms  $A^a$  cannot be chosen arbitrarily at this stage since they
need to satisfy (\ref{integrability}). However, what we can do is to
choose them still sufficiently general: Let them be of the form 
$A^a := \lambda^a \alpha$ where $\alpha$ is some arbitrary 1-form on $\S$
at $p$; then clearly $A^a \wedge A^b \equiv 0$ (at $p$) and the given
data at $p$ can be extended into some neighborhood of $p$ satisfying
(\ref{constraintsurface}). On the other hand, with $\lambda^a$ to be
free at our disposal, we can now indeed conclude (\ref{rhomorph}) from
(\ref{GJ}). 

Also note that at \emph{this} point the integrability conditions are
always satisfied, which in particular implies that at a given point in
$\S$ the 1-forms $A^a$ and the 2-forms $p_i$ can now be chosen
arbitrarily---still permitting choices for extensions of the fields
into a neighborhood of that point such that (\ref{constraintsurface})
holds true (cf.~eqs.~(\ref{G}) and (\ref{J})). In particular, this
implies that each line in (\ref{JJ}) has to vanish separately on the
constraint surface. In fact, the first two lines vanish by themselves
already, and the third one reproduces just (\ref{rhomorph})---at least
if we use that $C_{abc}$ is completely antisymmetric in its three
indices, which in fact is  reinforced in the fourth line of
(\ref{JJ}). 

Here some remark is in order: In the action that we used to derive the
Hamiltonian system the coefficients $C_{abc}$ entered already
completely antisymmetrically. Still, the constraints (\ref{G}) and
(\ref{J}) make sense also when $C^a_{bc}$ is antisymmetric in the last
two indices only. We performed the ensuing calculation in this relaxed
setting. Then we find that the first class property enforces the
antisymmetry in the first two indices as well, cf.~the fourth line of
(\ref{JJ}), i.e.~thus in all three indices. This is analogous to the
situation in the Poisson sigma model: The constraints (\ref{constr1})
are meaningful already in the more general setting of a general
contravariant 2-tensor ${\cal P}^{ij}$. Also there the first class
property enforces both, the antisymmetry of ${\cal P}^{ij}$ as well as
the Jacobi identity. Both conditions there have a meaning in terms of
Dirac structures: the first being the condition of isotropy, the other
one an integrability condition (cf.~also \cite{AlekseevStrobl} for
further details on this relation). In the three dimensional setting,
there are two algebraic conditions of this kind now,
eq.~(\ref{rhorho}) as well as the antisymmetry condition, 
\be C_{abc} = - C_{bac} , \label{antisymm} \ee
as well as two integrability conditions, eq.~(\ref{rhomorph}) and 
\be \label{Jakobi}
       C^e_{ab}C^d_{ce}+C^d_{ab,c}+\mbox{cycl}(abc)=
        C_{cab,e}\eta^{ed}+\rho^{di}\rho^j_a\rho^k_b\rho^l_c 
H_{ijkl} \, ,\ee
enforced by the vanishing of the last line in (\ref{JJ}).

\section{Axioms of $H_4$-twisted Courant algebroids} 
\label{sec:final}

In this section we want to extract the coordinate independent
information contained in the structural identities obtained above. In
section \ref{3dsigma} we already discovered that the differential
geometric setting is a vector bundle $E$ over a base manifold $M$,
equipped with a nondegenerate bilinear pairing $\eta$, a bundle map
$\rho$, cf.~Eq.~(\ref{rho}), which we will call the anchor of $E$, and
a closed 4-form $H$ on $M$. The main task of this section is to give a
meaning to the structural functions $C^a_{bc}$ and the interplay of
all the structural functions as dictated by the identities found
above. 

Let us be guided by the special well-known case of the Chern Simons
theory. This is obtained from $M$ being a point, $H$ and $\rho$
correspondingly zero, and $(E,\eta)$ thus being just a vector space
equipped with a non-degenerate bilinear form. In this case, $C^a_{bc}$
correspond to structure constants of a Lie algebra---in accordance with
this, eq.~(\ref{Jakobi}) reduces to the Jacobi identity---and $\eta$ is
invariant w.r.t.~the adjoint transformations of this Lie algebra, as
expressed by eq.~(\ref{antisymm}).

The most near-at-hand generalization of the above scenario over a
point would be that $C^a_{bc}$ defines a product on the space of
sections $\Gamma(E)$ of the bundle $E \to M$. However, this is in
conflict with the transformation properties found in (\ref{Ctrafo})! 
Let $D^a_{bc}$ be structural functions of a product of sections,
i.e.~if $e_a$ is a basis of sections in $E$ and we denote the product
by a bracket, one has 
\be [e_a, e_b] = D^c_{ab} e_c \, .\label{D} \ee
With this definition it is clear that under a local change of basis 
\be \widetilde e_a = M_a^b e_b \label{change} \ee
the   first index of $D_{abc}  =   \eta_{ad} D^d_{bc}$ transforms in a
$C^\infty$-linear  fashion, i.e.~that   $\widetilde D_{abc}$   will  be
simply proportional  to $M_a^d$, a matrix  $M$ with a lower $a$--index
(while the other indices can  produce also derivatives of $M$-matrices
etc---cf.~eq.~(\ref{Dtrafo}) below  as a  possible realization of this
requirement). This is however \emph{not} the case for $C_{abc}$, as we
learn from eq.~(\ref{Ctrafo}).

In order to cure this deficiency of $C$ to define structural functions
of a product of sections, we want to make an ansatz using the other
structural quantities at hand:
\be D_{abc} = C_{abc} + \alpha \eta_{ab,c} + \beta \eta_{bc,a} + 
\gamma \eta_{ca,b} \, . \label{ansatz} \ee
We observe that $a$-derivatives of $\eta$ transform in the following
way\footnote{By definition of $a$-derivatives, cf.~eq.~(\ref{kommaa}),
one has $(fg)_{,a} =  f_{,a} g + f g_{,a}$. Note, however, that such
type of derivatives do not commute. Instead, as a consequence of
(\ref{rhomorph}), one finds $f_{,ab} \equiv (f_{,a})_{,b} =
f_{,ba} +C^c_{ba} f_{,c}$.}
\be \widetilde \eta_{ab,c} = M_a^d M_b^e M_c^f  \eta_{de,f} + 
M_{a,d}^f M_b^e M_c^d \eta_{ef} + M_{b,d}^f M_a^e M_c^d \eta_{ef} \,
. \ee
Writing out the six terms coming from the antisymmetrization of the
second term in (\ref{Ctrafo}), it is now easy to see, that the
required $C^\infty$-linearity implies $\gamma = - \alpha =
\frac{1}{2}$, leaving $\beta$ arbitrary at this point. With such a 
choice of constants, $D$ thus defines a product by means of 
(\ref{D}). 

To fix the remaining constant, we regard the generalization of the
ad-invariance condition for $\eta$. For this purpose we first express
$\eta([e_a,e_b],e_c) + \eta(e_b,[e_a,e_c])$ in terms of the structural
functions $D$; using (\ref{D}), this becomes identical to $D_{cab} +
D_{bac}$. So it is the symmetrization over the first and the
\emph{third} index of $D$ (at this point it is not clear that $D$ will
define an antisymmetric product---and in fact it will not---in which
case one would be able to trade this into a symmetrization of the
first and second index, as one is used to from Lie algebras, cf.~also
Eq.~(\ref{antisymm})). Using, on the other hand, (\ref{ansatz}) with
the above choice for $\alpha$ and $\gamma$, we find
\be D_{abc} + D_{cba} = \eta_{ac,b} + (\beta - \2)
\left(\eta_{ab,c} + \eta_{cb,a} \right)\, , \label{hallo}
\ee
since $C_{abc}$ is completely antisymmetric
as entering the action (\ref{action2}).\footnote{There exists a more
involved argument using only (\ref{antisymm}) in orthonormal frames
and the transformation properties of the coefficients to arbitrary
frames to arrive at this conclusion from milder assumptions on
$C^c_{ab}$ (as described at the end of the previous section). However,
up to this point within this section all the argumentation can be done
already at the level of the action. The Hamiltonian perspective will
then be used only to extract the Jakobi condition.} While the first
term on the r.h.s.~of (\ref{hallo}) fits an ad-invariance condition
very well, the other terms are disturbing in this context. It is thus
comforting to see that they can be made to vanish by a unique choice
of the still free constant $\beta$ in our ansatz. Thus we are lead
to\footnote{The formal analogy of the expression for the difference
between $C$ and $D$ with the standard formula for a torsion-free,
metrical connection in a holonomic frame is somewhat striking at this
point.}
\be D_{abc} = C_{abc} + \2 \left(  \eta_{bc,a} + 
\eta_{ac,b} -\eta_{ab,c}  \right) \, . \label{ansatz2} \ee
Under arbitrary changes (\ref{change}) of frames, these coefficients
transform according to  
\be  \widetilde D_{abc} = M_a^d M_b^e M_c^f  D_{def} + M_a^d
\left(\eta_{de} M^e_{c,f} M^f_b-\eta_{de} M^e_{b,f} M^f_c + 
\eta_{ef} M_{b,d}^e  M_c^f  \right) \,
. \label{Dtrafo} \ee

We collect what we obtained up to now---it is already quite a lot, and
all this is coming from the action functional and its transformation
properties only: We have a vector bundle $E$ over $M$ together with an
anchor map $\rho \colon E \to TM$.
$E$ is equipped with a fiber metric $\eta$ and a product $[ \cdot,
\cdot ]$ on its sections. This product is \emph{not}
antisymmetric. Rather, according to (\ref{ansatz2}), we see that 
\be \eta(\psi_1, [\psi_2, \psi_3] +  [\psi_3, \psi_2]) =
\rho(\psi_1) \eta(\psi_2,\psi_3) \, , \label{Symmm} \ee
where $\psi_i$ are arbitrary sections of $E$ and $\rho(\psi_1)$ is the
vector field $\psi_1^a \rho_a^i \partial_i$. (This follows from
eq.~(\ref{ansatz2}) as follows: In the case that all three sections
are linearly independent, we can use them as part of a basis $e_a$. By
construction, (\ref{ansatz2}) holds in arbitrary
frames. Symmetrization over the last two indices in
(\ref{ansatz2}) indeed yields $D_{abc} + D_{acb} = \eta_{bc,a} \equiv
\rho_a^i \eta_{bc,i}$, which gives (\ref{Symmm}) for this case. Validity
of that equation in degenerate cases of linear dependence now follows
for example by continuity.) Since $\eta$ is non-degenerate, this
equation determines the symmetric part of the bracket uniquely. In a
completely analogous manner we conclude from (\ref{ansatz2})
(cf.~eq.~(\ref{hallo}) for $\beta=\frac{1}{2}$) the ad-invariance
condition of the fiber metric w.r.t.~the bracket on sections,
\be \eta([\psi_1,\psi_2],\psi_3) + \eta(\psi_2,[\psi_1,\psi_3]) =
\rho(\psi_1) \eta(\psi_2,\psi_3) \, . \label{adin} \ee
Note that the r.h.s.~of the last two equations is identical. Thus,
using a standard polarization argument ($\eta$ being symmetric), we
can rewrite these two equations according to
\be \eta([\psi' ,\psi], \psi ) = \2 \rho(\psi')
\eta(\psi,\psi) =   \eta(\psi' ,[\psi, \psi] ) \, , \label{drei} 
\ee
valid for arbitrary two sections $\psi$, $\psi'$ of $E$. 

There is still one further important property of the bracket that one
can conclude from the above definitions and transformation
properties. It concerns the relation of $[\psi_1,f \psi_2]$ to $f
[\psi_1,\psi_2]$, where $f$ is an arbitrary function on $M$. Let us
for this purpose choose $\psi_1$ and $\psi_2$ as the first two basis
elements of a local frame $e_a$ (we assume them to be linearly
independent and again conclude on the case of proportional sections by
continuity) and consider a change of frame (\ref{change}) with
$M_a^b=\delta_a^b$ if $(a,b) \neq (2,2)$ and $M_2^2 = f$ (we assume
$f$ to be nonzero, at least in a neighborhood of our
interest---otherwise $[\psi_1,f \psi_2]$ vanishes already by
bilinearity of the bracket). Then (\ref{Dtrafo}) yields (for $a \neq 2$)
\be \eta(e_a, [\psi_1,f \psi_2]) = \widetilde D_{a12} = f D_{a12} + 
\eta(e_a , \psi_2) \rho(\psi_1)f \, , \ee
since $M_1^e = \delta_1^e$ is constant and its derivative gives no
contribution. Thus we find the following Leibniz property of the
bracket: 
\be [\psi_1,f \psi_2] = f [\psi_1, \psi_2] +
\left(\rho(\psi_1)f\right) \psi_2 \, . \label{Leibniz} \ee

For later use we finally mention that (\ref{Symmm}) (or, equivalently,
the second equality of (\ref{drei})), can be also rewritten according
to \be [\psi,\psi] = \2 \rho^* \md \eta(\psi,\psi) \, , \label{psipsi}\ee
where, by definition, $\rho^*$ of some 1-form $\alpha=\alpha_i \md
x^i$ is just $
\alpha_i\rho^{ai} e_a$ (it is the fiberwise transpose of $\rho$ with a
subsequent use of $\eta$ to identify $E^*$ with $E$). From this and
(\ref{Leibniz}) one may also conclude for example about the behavior
of the bracket under multiplication of the first section w.r.t.~a
function:
\be  [f\psi_1,\psi_2] = f [\psi_1, \psi_2] -
\left(\rho(\psi_2)f\right) \psi_1 + \eta(\psi_1,\psi_2)\,\rho^*(\md f) 
 \, . \label{leftLeibniz} \ee
This also puts us in the position to express the general product or
bracket of two sections by means of the structural functions:
 \be [\psi_1,\psi_2] = \left(\psi_1^b
\psi_2^c D^a_{bc} + \rho(\psi_1) \psi_2^a - \rho(\psi_2) \psi_1^a +
\rho^{ai}(\psi_1^b)_{,i} (\psi_2)_b \right) e_a \, , \label{12}\ee 
where, certainly, 
$\rho(\psi_1) \psi_2^a = \rho^i_b \psi_1^b (\psi_2^a)_{,i}$ and
$\rho^{ai} = \eta^{ab} \rho^i_b$.

We now turn to the structural identities that we obtained in the
previous section. Here we need to emphasize that they were obtained in
an {\emph{orthonormal}} frame (or at least a frame where $\eta_{ab}$
is constant). Clearly, terms of importance in a general frame may be
absent in such a frame. One example is the Ad-invariance
condition of the metric tensor $\eta$: The condition (\ref{adin})
becomes (cf.~eq.~(\ref{hallo}) for $\beta=\frac{1}{2}$)
\be D_{abc} + D_{cba} = \eta_{ac,b} \label{Beispiel}\ee 
in an arbitrary local frame. Clearly the r.h.s.~of this equation
vanishes in an orthonormal frame and it is the question how one can
recover it from knowing the condition in orthonormal frames only.
On the other hand, we took great effort to derive transformation
properties of all structural functions with respect to \emph{general}
changes of a frame bundle basis (\ref{change}). Thus, we may proceed
as follows in principle: We note that within an \emph{orthonormal}
frame $D_{abc}= C_{abc}$. Thus we can replace in all of the identities
obtained in the previous section the structural functions $C$ by $D$
everywhere. Then we can apply the transformation formulas such as
(\ref{Dtrafo}) to all these identities, transforming them to a general
frame.

Let us illustrate this at the example of (\ref{antisymm}): Let us
assume that the frame $e_a$ is orthonormal, thus $\eta_{ab}$ in
particular constant, and $\widetilde e_a$ an arbitrary frame, so that
$\widetilde \eta_{ab}$ as given by eq.~(\ref{etatilde}) is in general
non-constant. Using the transformation property (\ref{Dtrafo}) we now 
compute 
\be  \widetilde D_{abc} + \widetilde D_{cba}  = 
M_a^d M_b^e M_c^f  (D_{def} + D_{fed})  + 
M_a^d \eta_{de} M^e_{c,f} M^f_b + M_c^d \eta_{de} M^e_{a,f} M^f_b \, ,
\label{Dtrafo2} \ee
where we made use of the fact that the last two terms in
(\ref{Dtrafo}) give no contribution when symmetrized over indices $a$
and $c$. In the orthonormal frame $e_a$ we have 
$D_{def} + D_{fed}= C_{def} + C_{fed} = - (C_{dfe} + C_{fde})$, which
\emph{vanishes} due 
to (\ref{antisymm}). On the other hand, the remaining two terms on the
r.h.s.~of (\ref{Dtrafo2}) combine into $\left(\widetilde
\eta_{ac}\right)_{,f} M^f_b$, which is nothing but
$\rho(\widetilde e_b) \widetilde
\eta_{ac}$. Thus indeed from (\ref{antisymm}) and the transformation
property (\ref{Dtrafo}) we find 
\be  \widetilde D_{abc} + \widetilde D_{cba}  = 
\rho(\widetilde e_b) \widetilde
\eta_{ac} \, , \label{tildesymm}\ee
i.e.~eq.~(\ref{Beispiel}) as it is to hold in an arbitrary frame. 

Thus we now could apply the same strategy on the other equations
obtained in the previous sections, such as for example to
(\ref{Jakobi}). Using (\ref{Dtrafo}) we would find, after quite a
lengthy calculation and on use of the other structural identities,
that, miraculously, (\ref{Jakobi}) would take the
\emph{same} form in an arbitrary frame (this certainly is partially
due to the fact, how we presented that formula---it certainly could be
rewritten in several inequivalent ways for constant metric
coefficients $\eta_{ab}$ such that this property holds no more true).
On the other hand, if we use e.g.~the \emph{transformation property} that one
\emph{obtains} upon choosing $\beta = 0$ in (\ref{ansatz}), i.e.~for 
\be E_{abc} = C_{abc} + \2
\left(\eta_{ac,b} -\eta_{ab,c}  \right) \, , \label{ansatz3} \ee
which one might use as coefficients of another product as we found
above, one would find (\ref{Jakobi}) to become more complicated in an
arbitrary frame. (Again one would have $E_{abc}=C_{abc}$ in an
orthonormal frame, could thus replace all $C$s by $E$s in
(\ref{Jakobi}), but now the $E$s would transform in a different way
than the $D$s, eq.~(\ref{Dtrafo}), which now would produce extra terms
similarly to what happened in the transition from (\ref{antisymm}) to
(\ref{tildesymm}) above. Using $D=C$ in orthonormal frames and the
transformation (\ref{Dtrafo}) on the other hand, will leave the
equation form-invariant, in fact upon usage of the other identities
obtained in the previous section). This observation may be used as
another argument besides (\ref{hallo}) for the choice
$\beta=\frac{1}{2}$ in the definition of the bracket.

There is, however, a more direct route to arrive at the missing axioms
as induced from the previous section. Before turning to it, but also
in preparation for it, let us briefly reconsider the relation of the
three different quantities $C_{abc}$, $D_{abc}$, and $E_{abc}$ from a
slightly more abstract perspective. First of all, we observe that
according to its definition in (\ref{ansatz3}), $E_{abc} =D_{a[bc]}$,
so $E$s are nothing but the structural functions of the
\emph{antisymmetrization} of the product (\ref{D}). So, if we denote
by $[[ \cdot, \cdot ]]$ the bracket defined via $E_{abc}$, i.e.~
\be [[ e_a , e_b ]] = E^c_{ab} e_c \ee
one has
\be [[\psi_1, \psi_2 ]] = 
\2 \left([\psi_1, \psi_2 ] - [\psi_2, \psi_1 ] \right) \, . 
\ee
This bracket is, by construction, antisymmetric, but, as mentioned
already, its other properties are slightly more involved than those
for the bracket $[
\cdot, \cdot ]$---like e.g.~instead of (\ref{adin}) one finds
\be \eta([[\psi_1,\psi_2]],\psi_3) + \eta(\psi_2,[[\psi_1,\psi_3]]) =
\rho(\psi_1) \eta(\psi_2,\psi_3) - \2\rho(\psi_2)\eta(\psi_3,\psi_1) -
\2\rho(\psi_3)\eta(\psi_1,\psi_2) \, , \label{adin2} \ee
for which reason we prefer to work with the previously introduced
non-antisymmetric bracket. Finally, according to its definition, what
is the relation of the coefficients $C_{abc}$ with the bracket? As
mentioned, $C_{abc}$ are \emph{not} the structure functions of any
product of sections. However, as we see from the very definition of
$D_{abc}$ in (\ref{D}), one has $C_{abc} = D_{[abc]}$. This implies
that if one defines
\be C(\psi_1,\psi_2,\psi_3) : = \frac{1}{6}\sum_{\sigma \in S_3} (-1)^{|\sigma|} 
\eta(\psi_{\sigma_1}, [ \psi_{\sigma_2}, \psi_{\sigma_3} ]) = 
\tfrac{1}{3} \eta(\psi_{1}, [[ \psi_{2}, \psi_{3} ]])
+ \mathrm{cycl}(123) \, , \label{C}\ee where $S_3$ denotes the
permutation group of three elements and $|\sigma|$ the parity of the
permutation element $\sigma$, we have $C(e_a,e_b,e_c) \equiv
C_{abc}$. So, eq.~(\ref{C}) relates $C$ in an arbitrary frame or as an
abstract object to the other two brackets and the scalar
product. Again, we remark that there is no way to induce a product
from $C$, in contrast to $D$ or $E$. (For example,
$E(\psi_1,\psi_2,\psi_3) = \eta(\psi_1, [[\psi_2, \psi_3]])$ is, in
contrast to (\ref{C}), $C^\infty$-linear in $\psi_1$, which thus
permits to define the product $[[ \cdot , \cdot ]]$ on sections of the
vector bundle $E$ from it).

We now come to the frame independent, abstract formulation of the
information contained in the three conditions (\ref{rhorho}),
(\ref{rhomorph}), (\ref{Jakobi}). Clearly, in more abstract terms, 
(\ref{rhorho}) just states that 
\be \rho \circ \rho^* = 0 \, , \label{rhorho*}\ee
where $\rho \colon E \to TM$ was the anchor map and $\rho^* \colon
T^*M\to E$ essentially its transpose, as introduced above. Here we
used that $\rho^i_a$ and $\eta_{ab}$ have a tensorial transformation
property, so that (\ref{rhorho}) in orthonormal frames applies the
likewise formula in arbitrary frames. Next we turn to
(\ref{rhomorph}). Also this equation is not difficult to
reinterpret. Let us for this purpose apply the map $\rho$ to
eq.~(\ref{12}):
\be \rho([\psi_1,\psi_2]) = \left(\psi_1^b
\psi_2^c D^a_{bc} + \rho(\psi_1) \psi_2^a - \rho(\psi_2)
\psi_1^a\right) \rho_a^i \partial_i \, , \label{rho12}\ee 
where we have already made use of (\ref{rhorho*}) to get rid of the
last term in (\ref{12}). This equation holds true in \emph{any}
frame. Thus also in an orthonormal frame, where we can replace $D$ by
$C$ and then make use of eq.~(\ref{rhomorph}), yielding---in this
orthonormal frame---: 
\be  \rho([\psi_1,\psi_2]) = \psi_1^b
\psi_2^c \left(\rho^j_{b}\rho^i_{c,j}- \rho^j_{c}\rho^i_{b,j}\right)\partial_i 
 + \rho(\psi_1) \psi_2^a \partial_i - \rho(\psi_2)
\psi_1^a \rho_a^i \partial_i \, . \ee
The r.h.s.~is, however, nothing but the commutator of the vector
fields $\rho(\psi_1)$ with $\rho(\psi_2)$. Thus we obtain, for an
\emph{arbitrary} choice of $\psi_1$, $\psi_2$ in $\Gamma(E)$, 
\be  \rho([\psi_1,\psi_2]) = [\rho(\psi_1),\rho(\psi_2)] \, . 
\label{rhomorphschoen}\ee
Note that here we only had to use an orthonormal frame as an
intermediary step. The resulting equation does no more show any
dependence on the frame; it is obviously sufficient and necessary to
guarantee (\ref{rhomorph}) in view of our definition of the
bracket---taking (\ref{rhorho*}) for granted! In fact, we can even
deduce (\ref{rhorho*}) from (\ref{rhomorphschoen}): Setting $\psi_1 =
\psi_2 $ and using (\ref{psipsi}), we find (\ref{rhorho*}) upon noting
that $\psi_1$ can be chosen such that $\md \eta(\psi_1,\psi_1)$ takes
any possible value at a given point. 

We will encounter a likewise fact in what follows next: the equation
that we will extract from (\ref{Jakobi}) will entail both,
eq.~(\ref{rhomorphschoen}) and (\ref{rhorho*}). Certainly, such facts
are true only upon usage of the Leibniz rules (\ref{Leibniz}) and
(\ref{leftLeibniz}), which we derived from the general transformation
and symmetry properties of $D_{abc}$ above. We now turn to the final,
most complicated condition, equation (\ref{Jakobi}). One may remark
also that it is the \emph{only} place where the 4-form $H$ enters
finally. 

To interpret eq.~(\ref{Jakobi}) within our present setting, we may
again remember to what it reduces for $M$ being a point, when it
becomes just the Jakobi identity for the Lie bracket. This may
motivate to consider the following expression:
\be J(\psi_1,\psi_2,\psi_3) := [\psi_1,[\psi_2,\psi_3]] - 
[[\psi_1,\psi_2],\psi_3]  - [\psi_2,[\psi_1,\psi_3]]  
\, . 
\ee
Note that certainly with the bracket $[\cdot , \cdot ]$ not being
antisymmetric, there are several inequivalent ways of writing the
Jakobiator. The above definition of $J$ corresponds to the choice
which measures the deviation of the bracket to satisfy a Leibniz
property with respect to itself, i.e.~that the adjoint transformation
$\mathrm{ad}_\psi : = [\psi , \cdot ]$ is a derivation of the bracket.  

To relate (\ref{Jakobi}) to $J$, we compute $J_{abc}:=J(e_a,e_b,e_c)$
with $e_a$ being an \emph{orthonormal} basis. In these frames we have  $D_{abc} =
C_{abc}$; using  (\ref{D}),
(\ref{Leibniz}), and (\ref{leftLeibniz}), one then easily establishes
the equivalence of  (\ref{Jakobi}) with 
\be J_{abc} = \rho^{di}\rho^j_a\rho^k_b\rho^l_c 
H_{ijkl} e_d \, . \label{Jakobi2} \ee

To relate this expression to one in a general basis, we first make use
of eqs. (\ref{Leibniz})--(\ref{leftLeibniz}) to obtain
\ba\label{Jdrei} J(\psi_1,\psi_2, f \psi_3) &=& f J(\psi_1,\psi_2, \psi_3) + 
\left([\rho(\psi_1),\rho(\psi_2)] -  \rho([\psi_1,\psi_2]) \right)f \,
\psi_3 \\\label{Jzwei} J(\psi_1,f \psi_2, \psi_3) &=&  f J(\psi_1,\psi_2, \psi_3) + 
\left([\rho(\psi_1),\rho(\psi_3)] -  \rho([\psi_1,\psi_3]) \right)f \,
\psi_2 \nonumber \\ && +  \left[\rho(\psi_1) \eta(\psi_2,\psi_3) -
\eta([\psi_1,\psi_2],\psi_3) - \eta(\psi_2,[\psi_1,\psi_3])\right] \,  \rho^*(\md f)
\nonumber \\
&&  -\eta(\psi_2,\psi_3) [ \rho^*(\md f), \psi_1] 
\\\label{Jeins} J(f \psi_1,\psi_2, \psi_3) &=&  f J(\psi_1,\psi_2, \psi_3) + 
\left([\rho(\psi_2),\rho(\psi_3)] -  \rho([\psi_2,\psi_3]) \right)f \,
\psi_1  \nonumber \\ && - \rho^*(\md f)  \left[\rho(\psi_2) \eta(\psi_1,\psi_3) -
\eta([\psi_2,\psi_1],\psi_3) - \eta(\psi_1,[\psi_2,\psi_3])\right] \nonumber \\
&& +\eta(\psi_1,\psi_3)[\rho^*(\md f),\psi_2]
-\eta(\psi_1,\psi_2)[\rho^*(\md f),\psi_3]
 \, . 
\ea
Since we already have the identities (\ref{rhomorphschoen}) and
(\ref{adin}) at our disposal, we see that it is
sufficient to show that
\be [\rho^* \md f , \psi ] = 0 \, , \label{forproof} \ee
for any $f \in C^\infty(M)$ and $\psi \in \Gamma(E)$, to obtain
that $J$ is $C^\infty(M)$-linear in each of its entries, i.e.~that it
is a tensorial object, $J \in \Gamma((E^*)^{\otimes 3} \otimes
E)$. Since, on the other hand, the r.h.s.~of (\ref{Jakobi2}) is
constructed by means of purely tensorial objects, this equation then
can immediately be considered as one valid for arbitrary frames, or,
likewisely can be rewritten as
\be\label{Jakobi3} J(\psi_1, \psi_2, \psi_3) =  \rho^*\left[ 
H(\cdot, \rho(\psi_1), \rho(\psi_2), \rho(\psi_3)) \right] \, , \ee
valid for arbitrary sections $\psi_i \in \Gamma(E)$. 

It thus remains to prove (\ref{forproof}). We distinguish tow cases,
$\eta$ having indefinite or definite signature. Thus we first assume,
that $\eta$ has a indefinite signature. Let $f\in C^\infty(M)$ and
$\psi\in\Gamma(E)$ be arbitrary, but with $\eta(\psi,\psi)\neq 0$.
(The particular case with $\eta(\psi,\psi)=0$ in (\ref{forproof}) then
follows by continuity from those cases). We put
$\psi_1=\eta(\psi,\psi)^{-1}\psi$.  Because of eqs.~(\ref{Leibniz})
and (\ref{rhorho*}) it is sufficient to show (\ref{forproof}) for
$\psi_1$ instead of $\psi$. Now we choose $\psi_2,\psi_3\in\Gamma(E)$
orthogonal to $\psi_1$ such that
$\eta(\psi_2,\psi_2)=\eta(\psi_3,\psi_3)=0$ and
$\eta(\psi_2,\psi_3)=1$. (This is always possible if the rank of $E$
is not too small). Finally we complete $\psi_1,\psi_2,\psi_3$
by orthonormal sections to get a basis $e_a = \psi_a$ in
$\Gamma(E)$. In this frame all components $\eta_{ab}$ are constant (at
least locally). The same holds true, if we replace $\psi_2$ and
$\psi_3$ by $\widetilde\psi_2=f\psi_2$ and
$\widetilde\psi_3=f^{-1}\psi_3$ (the case $f\equiv 0$ is trivial),
yielding a new basis $\widetilde \psi_a$, where in fact then one
obviously has $\widetilde \eta_{ab}=\eta_{ab}$. Hence we find by
(\ref{Jdrei}) and (\ref{Jzwei})
\[ J(\psi_1,\widetilde\psi_2,\widetilde\psi_3)
=ff^{-1}J(\psi_1,\psi_2,\psi_3)+\eta(\psi_2,\widetilde\psi_3)[\rho^*(\md
f),\psi_1] =J(\psi_1,\psi_2,\psi_3)+f^{-1}[\rho^*(\md f),\psi_1]\,,\]
and on the other hand due to (\ref{Jakobi2}) (note that $\eta_{ab}=\widetilde
\eta_{ab}$ is constant so that this formula can be applied in \emph{both} frames)
\[J(\psi_1,\widetilde\psi_2,\widetilde\psi_3) =  \rho^*\left[ 
H(\cdot, \rho(\psi_1), \rho(f\psi_2), \rho(f^{-1}\psi_3))
\right]=J(\psi_1,\psi_2,\psi_3)\,.\] Together this proves
(\ref{forproof}).


The case of definite signature now either follows by a
complexification argument (we can consider the sigma model with
imaginary fields) or by completing $\psi_1$ to an arbitrary
orthonormal frame and considering now $\widetilde\psi_2=\psi_2\cos
f+\psi_3\sin f$ and $\widetilde\psi_3=-\psi_2\sin f+\psi_3\cos f$. The
details of this second approach as well as the remaining
cases---i.e.~if there are no three linearly independent sections---are
left as an exercise to the reader.

Now, finally, we are in the position to give a concise, abstract
definition of an $H_4$-twisted Courant algebroid:

\begin{defi} 
A Courant algebroid twisted by a closed 4-form $H$
is a vector bundle $E \to M$ with fiber metric $\eta$, a
bundle map $\rho \colon E
\to TM$, and a bilinear product $[ \cdot , \cdot ]$ on $\Gamma(E)$ such
that
\ba\label{axiom1}
[\psi_1,f \psi_2] &=& f [\psi_1, \psi_2] +
\left(\rho(\psi_1)f\right) \psi_2 \\
\label{axiom2}
\rho(\psi_1)\eta(\psi_2,\psi_3)& =&
\eta(\psi_1 ,[\psi_2, \psi_3]) + \eta(\psi_1 ,[\psi_3, \psi_2]) \\
\label{axiom3}
\ [\psi_1,[\psi_2,\psi_3]] 
& = & 
[[\psi_1,\psi_2],\psi_3]  +  [\psi_2,[\psi_1,\psi_3]]  
+  \rho^*\left[ 
H(\cdot, \rho(\psi_1), \rho(\psi_2), \rho(\psi_3)) \right] \, .
\ea
\end{defi}

Here we took care to provide a possible minimal set of axioms. Even in
the known case of an ordinary (i.e.~nontwisted) Courant algebroid,
this has not always been the case in the mathematical literature. It
is, however, a fact that reversing our considerations from before and
starting with the above definition, \emph{all} the equations of this
section can be recovered. For example (\ref{psipsi}) is equivalent to
(\ref{axiom2}), hence (\ref{leftLeibniz}) can be deduced as before,
and thus also eqs.~(\ref{Jdrei})--(\ref{Jeins}). Now
(\ref{rhomorphschoen}) is a consequence of (\ref{axiom3}) and
(\ref{Jdrei}) due to the tensorial behaviour of
$J(\psi_1,\psi_2,\psi_3)$, and (\ref{rhorho*}) follows from
(\ref{axiom2}) with $\psi_1=\rho^*(\alpha)$ and
(\ref{rhomorphschoen}). Furthermore, (\ref{forproof}) can be obtained
from (\ref{axiom3}) and (\ref{psipsi}) (note that
$\eta(\psi_1,\psi_2)$ can be an arbitrary function, even when $\psi_2$
is fixed), and finally (\ref{adin}) follows from (\ref{Jzwei}),
(\ref{rhomorphschoen}), and (\ref{forproof}). 

\vspace{3mm}

At the end we shall discuss a concrete realization of such a twisted
Courant algebroid. Let $M$ be an arbitrary manifold, and consider
$E=TM\oplus T^*M$. We define $\rho((u,\alpha))=u$,
\[ \eta((u,\alpha),(v,\beta))=\alpha(v)+\beta(u) \, , \] 
and
\[ \left[(u,\alpha),(v,\beta)\bigr]=\bigl([u,v]_{\text{Lie}}\,,
	L_u\beta- L_v\alpha+\md(\alpha(v))+h(u,v,\cdot)\right)\]
for some arbitrary 3-form $h$. This implies
$\rho^*(\alpha)=(0,\alpha)$, and by a  calculation recommended to
the reader as an exercise one arrives at
\[J\left((u,\alpha),(v,\beta),(w,\gamma)\right)
	=\left(0,(\md h)(u,v,w,\cdot)\right)\,.\] 
So for the case that the 3-form $h$ is closed, one has an example of
an ordinary Courant algebroid. In fact, this is just the split exact
Courant algebroid mentioned in the Introduction ($h$ being the closed
3-form mentioned in footnote \ref{splitexact} in particular and its
deRham cohomology class is the Severa class which uniquely
characterizes an exact Courant algebroid \cite{SevLetts}). 

If, on the other hand, we consider the above data for an arbitrary
3-form $h$, we find an example of an $H$-twisted Courant algebroid
where the 4-form is simply $H=\md h$. (It is easy to verify that all
the axioms in Definition 1 hold true in this case). This however
implies that this example is one with an exact 4-form $H$ only. 

Such as twisted Poisson structures are best understood in terms of
appropriate substructures in (split) exact Courant algebroids,
$H$-twisted Courant structures can be understood as substructures of
the next higher analogue of these kind of nested structures
(particular degree three NPQ-manifolds in the corresponding language
mentioned briefly in the Introduction). Moreover, in the twisted
Poisson case it is only the cohomology class of the closed 3-form that
plays an inherent role from that perspective, exact 3-forms can be
``gauged away'' by a change of the splitting (cf., e.g.,
\cite{DSM}). We expect a likewise feature for the 4-form $H$ above
within the one step higher analogue, so that it may be worthwhile to
search for examples of $H$-twisted Courant structures with nonexact
4-forms also. 

\section*{Acknowledgements}
We are grateful to M.~Gr\"utzmann for discussions and to the 
Erwin Schr\"odinger Institute in Vienna for hospitality in the context
of the program ``Poisson sigma models, deformations, Lie algebroids,
and higher analogues'' in 2007.

\end{document}